\documentclass[prd,aps,floats,floatfix,superscriptaddress,preprintnumbers,
showpacs,eqsecnum,nofootinbib,twocolumn]{revtex4}
\usepackage{latexsym,array,theorem,mathrsfs,subfigure,bm,float}

\usepackage{psfrag}
\usepackage{amsfonts,amsmath,amssymb,latexsym,array,afterpage,
theorem,mathrsfs,bm,float,epsfig,color,graphicx,tabularx,slashbox,here,multirow}

\newcommand{\nn}{\nonumber \\}
\newcommand{\bea}{\begin{eqnarray}}
\newcommand{\ena}{\end{eqnarray}}
\newcommand{\beann}{\begin{eqnarray*}}
\newcommand{\enann}{\end{eqnarray*}}

\newcommand{\RB}{\overset{\scriptsize  (0)}{R}}

\newcommand{\Rf}{\overset{\scriptsize  (1)}{R}}
\newcommand{\GB}{\overset{\scriptsize  (0)}{G}}

\newcommand{\cGB}{\overset{\scriptsize  (0)}{{\cal G}}}

\newcommand{\gB}{\overset{\scriptsize  (0)}{g}}

\newcommand{\fB}{\overset{\scriptsize  (0)}{f}}

\newcommand{\TB}{\overset{\scriptsize  (0)}{T}}
\newcommand{\cTB}{\overset{\scriptsize  (0)}{{\cal T}}}
\newcommand{\Tf}{\overset{\scriptsize  (1)}{T}}
\newcommand{\cTf}{\overset{\scriptsize  (1)}{{\cal T}}}

\newcommand{\Mf}{\overset{\scriptsize  (1)}{M}}

\newcommand{\BoxB}{\overset{ \scriptsize (0)}{\Box}}
\newcommand{\nablaB}{\overset{ \scriptsize (0)}{\nabla}}

\newcommand{\nablag}{\overset{ \scriptsize (g)}{\nabla}}
\newcommand{\nablaf}{\overset{ \scriptsize (f)}{\nabla}}

\begin{document}

\baselineskip=12pt

\title{Dark matter in ghost-free bigravity theory:
\\[,3em] From a galaxy scale to the universe}
\author{Katsuki \sc{Aoki}}
\email{katsuki-a12@gravity.phys.waseda.ac.jp}
\affiliation{
Department of Physics, Waseda University,
Shinjuku, Tokyo 169-8555, Japan
}

\author{Kei-ichi \sc{Maeda}}
\email{maeda@waseda.ac.jp}
\affiliation{
Department of Physics, Waseda University,
Shinjuku, Tokyo 169-8555, Japan
}

\date{\today}

\begin{abstract}
We study the origin of dark matter based on 
the ghost-free bigravity theory with twin matter fluids. 
The present cosmic acceleration can be explained by the existence of graviton mass,
while dark matter is required in several cosmological situations 
[the galactic missing mass, the cosmic structure formation and 
the standard big-bang scenario (the cosmological nucleosynthesis vs the 
CMB observation)]. 
Assuming that the Compton wavelength of the massive graviton is shorter than 
 a galactic scale, we show 
the bigravity theory can explain  dark matter by twin matter fluid 
as well as the cosmic acceleration 
by tuning appropriate coupling constants.

\end{abstract}



\maketitle

\section{Introduction}
Whether a graviton has a mass or not is one of the most fundamental 
issues in physics. 
In general relativity (GR), it is well-known that the graviton 
is a massless spin-2 particle.
However, Fierz and Pauli proposed a massive spin-2 particle theory,
which is known as a unique ghost-free linear massive gravity theory\cite{FP}.
The present experimental solid constraint 
on the graviton mass is $m<7.1\times 10^{-23}$ eV \cite{graviton-mass, Will}.
Although a simple non-linear extension of the Fierz-Pauli massive gravity
theory  contains instabilities  called the Boulware-Deser ghost \cite{BD},
it was shown that the special choice of the interaction term 
can exclude  such a ghost state by de Rham et al\cite{dRGT, deser}.
However, this theory cannot describe the flat Friedmann universe, 
if the fictitious metric for the St\"{u}ckelberg field is Minkowski's one.
One may consider an inhomogeneous metric or extend it to de Sitter metric.
When we discuss an curved fictitious geometry, it may be natural 
for it to be dynamical. 
In fact the de Rham-Gabadadze-Tolley (dRGT) massive gravity 
theory has been generalized to 
such a bigravity theory, which is still ghost-free.
It contains a massless spin-2 particle 
and a massive spin-2 particle \cite{HassanRosen}. 

A phenomenological motivation to consider such theories 
relates to the discovery of dark energy and dark matter.
The cosmological parameters are now determined very precisely
\cite{Planck}.
Although standard big-bang cosmology explains many observed data,
those observations reveal new unsolved mysteries in cosmology, i.e., 
dark energy and dark matter.
Dark energy, which is the origin of 
the current accelerated expansion of the Universe, is one
of the most mysterious problems in modern cosmology \cite{IaSN}.
The acceleration might be due to some unknown matter
with a strange equation of state, or
might be due to a modification of GR.
As for the ghost-free massive gravity or bigravity theory without dark energy,
 many studies addressed 
the possibility to explain cosmic acceleration by 
the ``mass" term
\cite{MassiveCosmo1, MassiveCosmo2, MassiveCosmo3, MassiveCosmo4, MassiveCosmo5,MassiveCosmo6, MassiveCosmo7, Cosmology1,Cosmology2, Cosmology3, growth, Cosmology4, Cosmology5, with_twin_matter}.

In contrast to the massive gravity theory, 
bigravity theories also have 
a possibility to explain the origin of dark matter
\cite{footnote1}.
It is because there are two types of matter field in 
a bigravity theory.
If a matter field  interact with both metrics
 \cite{Cosmology5, matter_coupling1, matter_coupling2}, 
it will violate the equivalence principle, which
must hold in very high accuracy \cite{Will, d_matter}.
Hence we have to discuss two different matter fields, 
which are decoupled each other and interact only 
through two metric interactions.
We then call them twin matter fields \cite{Bimond}.

In the previous paper\cite{with_twin_matter},
 we found that both dark matter and dark energy components
in Friedmann equation can be obtained by modification of 
gravitational theory in the ghost-free bigravity theory .
However, dark matter is required not only in the big bang scenario
but also in the cosmological structure formation 
and as dark matter halos existing around galaxies. 
This paper will show 
a possibility to explain the origin of dark matter
in such situations.

The bigravity theory includes GR with/without a
cosmological constant as a special case.
If both metric are proportional, which we call a homothetic solution,
the basic equations are reduced to
two sets of the Einstein equations with cosmological constants,
which originate from the interaction terms of two metrics
\cite{Anisotropic2}.
Although  two matters must satisfy a fine tuned condition
in a homothetic solution, such a solution is an attractor and 
is obtained asymptotically from more generic initial conditions
\cite{with_twin_matter}.

The linear perturbations around a homothetic solution 
are decomposed into two eigenstates: the massless and massive graviton modes.
Note that these are the mass eigenstates,
whereas they are mixed up in the physical frame described by 
two metrics.
That is, the massless and massive modes couple to both twin matter fluids.
As a result, the perturbations of our spacetime are described 
by the linear combinations of the massless and massive modes.
Our spacetime is affected by another one of twin matter fluids via
the massless and massive graviton modes, 
and then there is a possibility 
such that dark matter component is originated by another twin matter.
The purpose of this paper is to investigate such a possibility.
Since dark matter is required in many situations,
 we shall discuss three typical evidences of dark matter: 
the content of the Universe, a galactic halo
and the cosmic structure formation.

The paper is organized as follows. Introducing the 
ghost-free bigravity, we summarize the basic equations 
and present a homothetic solution 
in \S \ref{HR}.
In \S \ref{linearized}, 
we then perform the perturbations around a homothetic solution.
We show that dark matter can be 
obtained from another one of twin matter fluids 
from a galactic scale to a cosmological scale
in \S \ref{origin_dark_matter}. 
Assuming the Compton wavelength of the massive graviton
is shorter than a galactic scale, 
another twin matter can play a role of  dark matter 
in our world for all scales.
We summarize our results and give some remarks in \S 
\ref{summary}.
In Appendix A,
we evaluate  the values of the graviton mass 
and a cosmological constant for given coupling parameters.
We also present the basic equations for the gauge invariant perturbations
in a homothetic background solution.

\section{Bigravity Theory}\label{HR}
\subsection{Hassan-Rosen bigravity model}
In the present papers, we focus on the ghost-free 
bigravity theory proposed by Hassan and 
Rosen, which  action is given by
\begin{eqnarray}
S &=&\frac{1}{2 \kappa _g^2} \int d^4x \sqrt{-g}R(g)+ \frac{1}{2 \kappa _f^2}
 \int d^4x \sqrt{-f} \mathcal{R}(f) \nonumber \\
&+&
S^{[\text{m}]}(g,f, \psi_g, \psi_f)
-\frac{m^2}{ \kappa ^2} \int d^4x \sqrt{-g} \mathscr{U}(g,f) 
\,,
\label{action}
\end{eqnarray}
where $g_{\mu\nu}$ and $f_{\mu\nu}$ are two dynamical metrics, and
$R(g)$ and $\mathcal{R}(f)$ are those Ricci scalars, respectively.
  $\kappa_g^2=8\pi G$ and $\kappa_f^2=8\pi \mathcal{G}$ are 
the corresponding gravitational constants, 
while $\kappa$ is defined by $\kappa^2=\kappa_g^2+\kappa_f^2$. 
We assume that the matter action $S^{(\text{m})}$ 
is divided into two parts:
\bea
S^{(\text{m})}(g,f, \psi_g, \psi_f)
=S_g^{[\text{m}]}(g,\psi_g)+S_f^{[\text{m}]}(f,\psi_f)
\,,
\ena
i.e.,  matter fields  $\psi_g$ and $\psi_f$ are coupled only to the $g$-metric 
and to the $f$-metric, respectively.
This restriction guarantees  the weak equivalence principle.
We call the $g$-matter $\psi_g$  and the $f$-matter $\psi_f$ 
 twin matter fluids.

 The ghost-free interaction term between two metrics is given by
\begin{equation}
\mathscr{U}(g,f)=\sum^4_{k=0}b_k\mathscr{U}_k(\gamma)
\,,
\end{equation}
{\setlength\arraycolsep{2pt}\begin{eqnarray}
&&\mathscr{U}_0(\gamma)=-\frac{1}{4!}\epsilon_{\mu\nu\rho\sigma} 
\epsilon^{\mu\nu\rho\sigma}\,, \nonumber \\
&&\mathscr{U}_1(\gamma)=-\frac{1}{3!}\epsilon_{\mu\nu\rho\sigma} 
\epsilon^{\alpha\nu\rho\sigma}
{\gamma ^{\mu}}_{\alpha}\,, \nonumber \\
&&\mathscr{U}_2(\gamma)=-\frac{1}{4}\epsilon_{\mu\nu\rho\sigma} 
\epsilon^{\alpha\beta\rho\sigma}
{\gamma ^{\mu}}_{\alpha}{\gamma ^{\nu}}_{\beta}\,, \\
&&\mathscr{U}_3(\gamma)=-\frac{1}{3!}\epsilon_{\mu\nu\rho\sigma} 
\epsilon^{\alpha\beta\gamma\sigma}
{\gamma ^{\mu}}_{\alpha}{\gamma ^{\nu}}_{\beta}{\gamma ^{\rho}}_{\gamma}\,, 
\nonumber \\
&&\mathscr{U}_4(\gamma)=-\frac{1}{4!}\epsilon_{\mu\nu\rho\sigma} 
\epsilon^{\alpha\beta\gamma\delta}
{\gamma ^{\mu}}_{\alpha}{\gamma ^{\nu}}_{\beta}{\gamma ^{\rho}}_{\gamma}
{\gamma ^{\sigma}}_{\delta}\,,
\nonumber
\end{eqnarray}}
where $b_k$ are coupling constants, while ${\gamma^{\mu}}_{\nu}$ is 
defined by 
\begin{equation}
{\gamma^{\mu}}_{\rho}{\gamma^{\rho}}_{\nu}
=g^{\mu\rho}f_{\rho\nu}
\,. 
\label{gamma2_metric}
\end{equation}
In order to take the square root to obtain the explicit form of
 ${\gamma^{\mu}}_{\nu}$, we shall introduce the tetrad systems,
$\{e_\mu^{(a)}\}$ and $\{\omega_\mu^{(a)}\}$, 
which are defined by
\begin{equation}
g_{\mu\nu}=\eta_{ab}e_\mu^{(a)}e_\nu^{(b)}
\,,~~
f_{\mu\nu}=\eta_{ab}\omega_\mu^{(a)}\omega_\nu^{(b)}
\,,
\end{equation}
with an additional constraint 
$e^\mu{}_{(a)} \omega_{\mu (b)}
=e^\mu{}_{(b)} \omega_{\mu (a)}$.
This constraint guarantees that the tetrad description is 
equivalent to the metric description.

We then find 
\bea
{\gamma^{\mu}}_{\nu}=\epsilon \eta_{ab} e^{\mu}{}^{(a)}\omega_\nu^{(b)}
\,,
\ena
where $\epsilon=\pm 1$ comes from the square root. 
As for the directions of tetrads, we 
choose that $e_\mu^{(0)}dx^\mu$ and $\omega_\mu^{(0)}dx^\mu$ 
are future-directed for $dt>0$.
Changing the sign of $\epsilon$ corresponds to 
 the following transformation
\begin{equation}
{\gamma^{\mu}}_{\nu} \leftrightarrow  -{\gamma^{\mu}}_{\nu}
\,,
\end{equation}
for which the interaction term is invariant 
by changing the sign of 
the coupling constants as
\bea
b_k\leftrightarrow (-1)^k b_k ~~~(k=0-4)
\,.
\ena

Taking the variation of the action with respect to $g_{\mu\nu}$ and
$f_{\mu\nu}$, we find two sets of the Einstein equations:
{\setlength\arraycolsep{2pt}\begin{eqnarray}
{G ^{\mu}}_{\nu} &=&
\kappa _g^2 ( {T ^{ [\gamma ] \mu} }_{\nu} 
+ {T^{\text{[m]} \mu} }_{\nu} ) \label{g-equation}, \\
{ \mathcal{G} ^{\mu}}_{\nu} &=& \kappa _f^2
( {\mathcal{T} ^{ [\gamma ] \mu} }_{\nu} 
+ {\mathcal{T}^{\text{[m]} \mu} }_{\nu} ), \label{f-equation}
\end{eqnarray}}
where ${G ^{\mu}}_{\nu}$ and ${ \mathcal{G} ^{\mu}}_{\nu} $ are the Einstein 
tensors for $g_{\mu\nu}$ and $f_{\mu\nu}$, respectively. 
The matter energy-momentum tensors 
are given by 
\bea
{ T^{\text{[m]}}}_{\mu\nu} &=&-2{\delta S_g^{[\text{m}]}
\over \delta g^{\mu\nu}}
\nn
{\mathcal{T}^{\text{[m]}}}_{\mu\nu}&=&-2{\delta S_f^{[\text{m}]}
\over \delta f^{\mu\nu}} 
\,.
\ena
 The $\gamma$-``energy-momentum" tensors from the interaction term
 are given by
{\setlength\arraycolsep{2pt}\begin{eqnarray}
&&{T ^{ [\gamma ] \mu} }_{\nu}=\frac{m^2}{\kappa^2} \
 ({\tau ^{\mu}}_{\nu} - \mathscr{U} {\delta ^{\mu}}_{\nu} \label{T(g)} ), 
\\
&& {\mathcal{T} ^{ [\gamma ] \mu} }_{\nu}  = -\frac{\sqrt{-g}}{\sqrt{-f}} 
\frac{m^2}{\kappa^2} {\tau ^{\mu}}_{\nu} \label{T(f)},
\end{eqnarray}}
with
{\setlength\arraycolsep{2pt}\begin{eqnarray*}
\tau^\mu_{~\nu}&=&
\{b_1\,\mathscr{U}_0+b_2\,\mathscr{U}_1+b_3\,\mathscr{U}_2
+b_4\,\mathscr{U}_3\}\gamma^\mu_{~\nu} \notag 
\nonumber \\
&-&\{b_2\,\mathscr{U}_0+b_3\,\mathscr{U}_1+b_4\,
\mathscr{U}_2\}(\gamma^2)^\mu_{~\nu}  \\
&+&\{b_3\,\mathscr{U}_0+b_4\,\mathscr{U}_1\}(\gamma^3)^\mu_{~\nu} 
\notag  \\
&-&b_4\,\mathscr{U}_0\,(\gamma^4)^\mu_{~\nu}
\,.
\end{eqnarray*}}

The energy-momenta of matter fields are assumed to be 
conserved individually as
\begin{equation}
\overset{(g)}{\nabla} _{\mu}{T^{ [\text{m}] \mu} }_{\nu}=0\,,\;
\overset{(f)}{\nabla} _{\mu} {\mathcal{T} ^{ [\text{m} ] \mu} }_{\nu} =0
\,, 
\label{c1}
\end{equation}  
where $\overset{(g)}{\nabla} _{\mu}$ and $\overset{(f)}{\nabla} _{\mu}$ are 
covariant derivatives with respect to $g_{\mu\nu}$ and $f_{\mu\nu}$. 
From the contracted Bianchi identities for 
\eqref{g-equation} and \eqref{f-equation}, 
the conservation of  the $\gamma$-``energy-momenta"  is
also guaranteed as
\begin{equation}
\overset{(g)}{\nabla} _{\mu}{T ^{ [\gamma] \mu} }_{\nu}=0\,,\;
\overset{(f)}{\nabla} _{\mu} {\mathcal{T} ^{ [\gamma] \mu} }_{\nu} =0
\,.
\label{c2}
\end{equation}

\subsection{Homothetic solution}
\label{homothetic}
First we give one simple solution,
in which we assume that two metrics are proportional;
\bea
f_{\mu\nu}=K^2\, g_{\mu\nu}
\,,
\ena
where $K$ is a scalar function.
In this case, since we find the tensor $\gamma^\mu{}_\nu=K\,
\delta^\mu{}_\nu$,  the $\gamma$-``energy-momentum"  is given by
\begin{align*}
\kappa_g^2{T ^{ [\gamma ] \mu} }_{\nu} &=-\Lambda_g(K)\delta^\mu{}_\nu
\,,
\\
\kappa_f^2{\mathcal{T} ^{ [\gamma ] \mu} }_{\nu} &=-\Lambda_f(K)
\delta^\mu{}_\nu
\,,
\end{align*}
 where 
\begin{align}
\Lambda_g(K)&=m^2\frac{\kappa_g^2}{\kappa^2}\,
\left(b_0+3b_1 K+3b_2 K^2+b_3 K^3\right)\,,
\nn
\Lambda_f(K)&=m^2\frac{\kappa_f^2}{\kappa^2}
\,\left(b_4+3b_3 K^{-1}+3b_2 K^{-2}+b_1K^{-3} \right)
\,. \label{eff_cc}
\end{align}

From the energy-momentum conservation (\ref{c2}), 
we find that $K$ is a constant.
As a result, we find two sets of the Einstein
equations with cosmological constants
$\Lambda_g$ and $\Lambda_f$:
\bea
G_{\mu\nu}(g)+\Lambda_g\,g_{\mu\nu}&=&\kappa_g^2 {T^{\text{[m]}}}_{\mu\nu}\,,
\label{homothetic_g}
\\
\mathcal{G}_{\mu\nu}(f)+\Lambda_f\,f_{\mu\nu} &=& \kappa _f^2
 {\mathcal{T}^{\text{[m]} } }_{\mu\nu} 
\label{homothetic_f}
\,.
\ena
Since two metrics are proportional, we have the constraints on 
the cosmological constants and matter fields as
\bea
\Lambda_g(K)&=&K^2\Lambda_f(K)\,,
\label{eq_K}
\\
\kappa_f^2{\mathcal{T}^{\text{[m]} } }_{\mu\nu} 
&=&\kappa_g^2\,{T^{\text{[m]}}}_{\mu\nu}
\,.
\ena
The quartic equation (\ref{eq_K}) for $K$ has 
 at most four real roots, which give 
four different cosmological constants.
The basic equations  (\ref{homothetic_g}) 
(or (\ref{homothetic_f})) are just the Einstein equations 
in GR with a cosmological constant.
Hence any solutions in GR with a cosmological constant are always
the solutions in the present bigravity theory.
We shall call these solutions homothetic solutions because of
the proportionality of two metrics.

\section{Linearization of the  bigravity theory}
\label{linearized}
\subsection{The perturbations around a homothetic solution}
\label{perturbation_homothetic}
The bigravity theory contains both massless and massive spin-2 particles.
It becomes clear when we discuss the linear perturbations around a homothetic solution. 
Note that a homothetic solution is an attractor in a cosmological setting
\cite{with_twin_matter}.

The unperturbed solution is assumed to be  homothetic, i.e.,
\bea
\gB_{\mu\nu}~~~{\rm and}~~~  \fB_{\mu\nu}=K^2  \gB_{\mu\nu}
\,,
\ena
which is the solution of two Einstein equations:
\begin{align}
\GB{}^{\mu}_{~\nu}(\gB)&=-\Lambda_g(K) \delta^\mu_{~\nu}+
\kappa_g^2 \TB{}^{{\rm [m]}\,\mu}_{~~~~~\nu}  
\label{homo_basic_eqs1}
\,,
\\
\cGB{}^{\mu}_{~\nu}(\fB)&=-\Lambda_f(K) \delta^\mu_{~\nu}+
\kappa_f^2 \cTB{}^{{\rm [m]}\,\mu}_{~~~~~\nu}
 \,,       
\label{homo_basic_eqs2}
\end{align}
A constant $K$ is determined by the quartic equation (\ref{eq_K}), 
and the matter energy-momenta satisfy the following condition:
\bea
\kappa_f^2 
\cTB{}^{[m]\mu}_{~~~~~\nu}={1\over K^2 }
\kappa_g^2 
\TB{}^{[m]\mu}_{~~~~~\nu}
\,.
\ena

We then consider the following perturbations:
\begin{align}
g_{\mu\nu}&=\gB{}_{\mu\nu}+h^{[g]}_{\mu\nu},\\
f_{\mu\nu}&=\fB{}_{\mu\nu}+ K^2h^{[f]}_{\mu\nu}
=K^2\left(\gB{}_{\mu\nu}+h^{[f]}_{\mu\nu}\right)
\end{align}
where $|h^{[g]}_{\mu\nu}|, |h^{[f]}_{\mu\nu}| \ll |\gB{}_{\mu\nu}|$.
The suffixes of $h^{[g]}_{\mu\nu}$ 
as well as  $h^{[f]}_{\mu\nu}$ are raised and lowered by the background metric
$\gB_{\mu\nu}$.

The energy-momentum tensors of twin matter fluid 
and $\gamma$-``energy-momentum" ones from the interaction terms 
can be expanded as
\begin{align}
\kappa_g^2T^{[\rm m]\mu}{}_{\nu}
&=\kappa_g^2\left[\TB{}^{[\rm m]\mu}{}_{\nu}+ \Tf{}^{[\rm m]\mu}{}_{\nu}\right] 
,\\
K^2\kappa_f^2{\cal T}^{[\rm m]\mu}{}_{\nu}
&=K^2\kappa_f^2\left[\cTB{}^{[\rm m]\mu}{}_{\nu}+ \cTf{}^{[\rm m]\mu}{}_{\nu}\right] 
\end{align}
and 
\begin{align}
\kappa_g^2{T^{[\gamma]\mu}}_{\nu}
&=-\Lambda_g{\delta^{\mu}}_{\nu}
+ \frac{m_g^2}{2}({h^{[-]}{}^{\mu}}_{\nu}-h^{[-]}{\delta^{\mu}}_{\nu}), \\
K^2\kappa_f^2{T^{[\gamma]\mu}}_{\nu}
&=-K^2\Lambda_f{\delta^{\mu}}_{\nu}
- \frac{m_f^2}{2}({h^{[-]}{}^{\mu}}_{\nu}-h^{[-]}{\delta^{\mu}}_{\nu})
\,,
\end{align}
respectively, 
where
\begin{align}
m_g^2&:=\frac{m^2\kappa_g^2}{\kappa^2}(b_1K+2b_2K^2+b_3K^3), \\
m_f^2&:=\frac{m^2\kappa_f^2}{K^2\kappa^2}(b_1K+2b_2K^2+b_3K^3)
\,.
\end{align}
Here we have introduced new variables 
$h^{[-]}_{\mu\nu}$ and $h^{[+]}_{\mu\nu}$ from two metric perturbations
as 
\begin{align}
h^{[-]}_{\mu\nu}&=h^{[g]}_{\mu\nu}-h^{[f]}_{\mu\nu}
,\nn
h^{[+]}_{\mu\nu}&=\frac{m_f^2}{m_{\rm eff}^2}h^{[g]}_{\mu\nu}+
\frac{m_g^2}{m_{\rm eff}^2}h^{[f]}_{\mu\nu}
\end{align}
with 
\begin{align}
m_{\rm eff}^2&:=m_g^2+m_f^2 
\nn&=\frac{m^2}{\kappa^2}
\left(\kappa_g^2+\frac{\kappa_f^2}{K^2}\right)
(b_1K+2b_2K^2+b_3 K^3)
\,.\label{eff_mass}
\end{align}

The first order perturbation equations are then given by
\begin{align}
&\gB{}^{\mu\rho} \Rf{}_{\rho \nu}(h^{[+]})-\RB{}^{\rho(\mu}h^{[+]}_{\nu)\rho}
=\Mf{}^{[+]\mu}{}_{\nu} \label{massless_eq}
\,,
\\
&\gB{}^{\mu\rho} \Rf{}_{\rho \nu}(h^{[-]})-\RB{}^{\rho(\mu}h^{[-]}_{\nu)\rho}
\nn
&~~~
+\frac{m_{\rm eff}^2}{4}\left(2h^{[-]\mu}{}_{\nu}+h^{[-]}\delta^{\mu}{}_{\nu}\right)
=\Mf{}^{[-]\mu}{}_{\nu} \label{massive_eq}
\,,
\end{align}
where $\Rf{}_{\mu\nu}$ denotes the linearized
Ricci tensor,
which is defined for a metric perturbation $h_{\mu\nu}$ 
by 
\begin{align}
\Rf_{\mu\nu}(h):={1\over 2}\Biggl[
&- \nablaB_\mu \nablaB_\nu h
-\BoxB
h_{\mu\nu} \nn
&+ \nablaB{}^\alpha(
\nablaB_\nu h_{\alpha\mu}) +\nablaB{}^\alpha
(\nablaB_\mu h_{\alpha\nu}) \Biggl]
\,,
\end{align}
and 
the matter perturbations $\Mf {}^{[\pm]\mu}{}_{\nu}$ 
are defined by 
\begin{align}
\Mf {}^{[-]\mu}{}_{\nu}&:=
\kappa_g^2 \left[\Tf{}^{[\rm m]}{}^{\mu}{}_{\nu}
-\frac{1}{2}\Tf{}^{[\rm m]}\delta^{\mu}{}_{\nu}\right]
\nn &
-K^2 \kappa_f^2 \left[
\cTf {}^{[\rm m]}{}^{\mu}{}_{\nu}
-\frac{1}{2}\cTf{}^{[\rm m]}\delta^{\mu}{}_{\nu}\right]
, \nn
\Mf{}^{[+]\mu}{}_{\nu}&:=
\frac{m_f^2}{m_{\rm eff}^2}  \kappa_g^2\left[\Tf{}^{[\rm m]}{}^{\mu}{}_{\nu}
-\frac{1}{2}\Tf{}^{[\rm m]}\delta^{\mu}{}_{\nu}\right]
\nn 
&+\frac{m_g^2}{m_{\rm eff}^2}K^2\kappa_f^2 \left[
\cTf{}^{[\rm m]}{}^{\mu}{}_{\nu}
-\frac{1}{2}\cTf{}^{[m]}\delta^{\mu}{}_{\nu}\right]
\,,
\end{align}
which are linear combinations of $g$- and $f$-matter perturbations.
Eqs. (\ref{massless_eq}) and (\ref{massive_eq}) are decoupled,
and then they provide two mass eigenstates. We find that 
$h^{[+]}_{\mu\nu}$ and $h^{[-]}_{\mu\nu}$ describe massless 
and massive modes, respectively, and 
$m_{\rm eff}$ denotes a graviton mass of the massive mode in the homothetic 
background spacetime.

The Bianchi identity ($\nablag{}_\mu G^\mu{}_\nu=0$)
gives the conservation of $\gamma$-``energy-momentum" tensor,
i.e.,
\bea
\nablag{}_\mu T^{[\gamma]}{}^{\mu}{}_{\nu}=0
\,,
\ena
which perturbation gives 
the constraint on the massive mode $h^{[-]}_{\alpha\beta}$:
\begin{align}
\nablaB{}_\mu \left(\kappa_g^2 \Tf{}^{[\gamma]}{}^{\mu}{}_{\nu}\right)
=\frac{m_g^2}{2}\left[
-\nablaB{}_\mu h^{[-]\mu} {}_{\nu}+\nablaB{}_{\nu} h^{[-]}\right]=0
\,.
\end{align}
Since $m_{g}^2\neq 0$,
we find 
\bea
\nablaB{}_{\mu} h^{[-]\mu}{}_{\nu}=\nablaB{}_{\nu} h^{[-]}
\label{pert_varphi-Bianchi}
\,.
\ena
From another conservation equation
$\nablaf{}_{\mu} \mathcal{T}^{[\gamma]\mu}{}_{\nu}=0$
gives the same constraint equation.

Taking a trace of Eq. (\ref{massive_eq}) and using 
Eq. (\ref{pert_varphi-Bianchi}), we find 
\begin{align}
&(3 m_{\rm eff}^2-2\Lambda_g)
h^{[-]} \nn
&=
\kappa_g^2(2\TB{}^{[\rm m]}_{\alpha\beta} h^{[-]\alpha\beta}
-\TB{}^{[\rm m]}h^{[-]})+
2\overset{\tiny (1)}{M}{}^{[\rm -]}{}^\alpha_{~\alpha}
\,.
\label{pert_varphi-trace}
\end{align}
Eqs. (\ref{pert_varphi-Bianchi}) and (\ref{pert_varphi-trace}) 
give five constraint equations on $h^{[-]}_{\alpha\beta}$.
There is no gauge freedom because $h^{[-]}_{\alpha\beta}$ 
is a gauge invariant tensor. 
This is consistent with the fact that 
a massive graviton 
has five degrees of freedom. 

Using these constraints, we rewrite the above perturbation equations as
\begin{align}
&-\nablaB{}_\mu \nablaB{}_\nu \, h^{[+]} 
-\BoxB  h^{[+]}_{\mu \nu} 
+ 2\nablaB{}_{(\nu }\Big[
\nablaB{}^\alpha \, h^{[+]}_{ \mu)\alpha } \Big] 
\nn
&~~~~~~~~~
-2\RB{}_{\mu}{}^\alpha{}_{\nu}{}^\beta h^{[+]}_{\alpha\beta}
=
2\Mf{}^{[+]}_{\mu \nu}
\,,
\\
&-\nablaB{}_{\mu} \nablaB{}_{\nu}\, h^{[-]}
-\BoxB h^{[-]}_{\mu \nu}
-2\RB{}_{\mu}{}^\alpha{}_{\nu}{}^\beta h^{[-]}_{\alpha\beta}
\nn
&~~~~~~~~~
+m_{\rm eff}^2
\Big(h^{[-]}_{\mu\nu}+{1\over 2}h^{[-]}\gB{}_{\mu\nu}\Big)
=
2\Mf{}^{[-]}_{\mu\nu}
\,,
\label{pert_varphi}
\end{align}
where we have used 
\begin{align}
\nablaB{}^\alpha(
\nablaB_\nu \chi_{\mu\alpha}) 
&=\nablaB_\nu(
 \nablaB{}^\alpha\chi_{\mu\alpha}) 
+\RB{}^{~~\alpha\beta}_{\mu~~~~\nu}\chi_{\alpha\beta}
+\RB{}^{\rho}_{~\nu}\chi_{\mu\rho}
\,.
\end{align}

Since two modes are decoupled, 
we shall analyze them separately, 
and then discuss the physical perturbations in the $g$- and $f$-worlds,
which  are 
represented as
\begin{align}
h^{[g]}_{\mu\nu}
&=
h^{[+]}_{\mu\nu}+\frac{m_g^2}{m_{\rm eff}^2}h^{[-]}_{\mu\nu}
, \nn
h^{[f]}_{\mu\nu}
&=
h^{[+]}_{\mu\nu}-\frac{m_f^2}{m_{\rm eff}^2}h^{[-]}_{\mu\nu}\,.
\label{g_f_perturbation}
\end{align}

Since the massive mode $h^{[-]}_{\mu\nu}$ does not propagate beyond 
the scale of the Compton wavelength of a massive graviton,
the spacetime perturbations are dominated by
the massless mode $h^{[+]}_{\mu\nu}$ 
in a large scale system beyond the Compton wavelength.
The massless mode couples to both twin matter fluids.
As a result, 
there exists a possibility that
 the $f$-matter fluid
behaves like a dark matter component in $g$-world 
via a massless graviton mode, which we will discuss 
in what follows.

\section{The origin of dark matter}
\label{origin_dark_matter}
In this section, we will analyze whether the $f$-matter field can be 
dark matter in our $g$-world. 
We believe from observation that the evidence of dark matter 
appears in three situations; (A) dark matter in the Friedmann equation, (B) 
a dark halo at 
a galaxy scale, and (C) CDM in cosmic structure formation.
So we discuss them in order.
\subsection{Cosmic pie}
\label{sec_cosmic_pie}
First we discuss the pie chart of the content of the Universe.
The amount of dark matter is about 5 times as large as the baryonic matter.
Since we discussed the details of dynamics of the FLRW spacetime and 
 possibility to explain the dark matter component 
by the $f$-matter fluid in \cite{with_twin_matter},
we give a brief overview here.

In order to explain the cosmic pie, 
we consider the homogeneous and isotropic spacetime, 
which metrics are given by
\begin{align}
ds_g^2=-N_g^2dt^2+a_g^2\left(\frac{dr^2}{1-kr^2} +r^2 d\Omega ^2\right),\\
ds_f^2=-N_f^2dt^2+a_f^2\left(\frac{dr^2}{1-kr^2} +r^2 d\Omega ^2\right),
\end{align}
where $N_g$ and $N_f$ are lapse function, while 
$a_g$ and $a_f$ are scale factors for $g_{\mu\nu}$ and $f_{\mu\nu}$,
respectively. Using the gauge freedom, we can set $N_g=1$ without loss of generality.

For generic initial data, 
the ratios $N_f/N_g$ and $a_f/a_g$ can approach to the same constant $K$ 
 given by \eqref{eq_K},
as the universe expands, i.e.,
the homothetic solution is  an attractor in the present system.
The dynamical time scale is about $m_{\rm eff}^{-1}$. 
As a result, near the attractor, i.e., near the present stage of the universe, 
 the evolution of the  universe is described 
by the effective Freidmann equation
\begin{align}
H_g^2+\frac{k}{a_g^2}=\frac{\Lambda_g}{3}+\frac{\kappa_{\rm eff}^2}{ 3}
\left[\rho_g+\rho_{\rm D}\right] \, ,
\end{align}
where
\begin{align}
\kappa_{\rm eff}^2&=
\kappa_g^2
\left[1-\frac{3m_g^2}{3m_{\rm eff}^2-2\Lambda_g}\right] \,,
\end{align}
\begin{align}
\rho_{\rm D}&=\frac{3m_f^2}{3m_f^2-2\Lambda_g}
\,K^4\rho_f\,,
\end{align}
and $\rho_{g}$ and $\rho_f$ are energy densities of $g$- and $f$-matter, respectively
\cite{with_twin_matter}.
$H_g=\dot{a}_g/a_g$ is the Hubble parameter where
a dot denotes the derivative with respect to $t$.
$\kappa_{\rm eff}^2$ is the effective gravitational constant, and
$\rho_{\rm D}$ is regarded as the energy density of a dark component in 
the $g$-world, i.e., another one of twin matter fluids works as dark matter 
through the interaction term between two metrics.

If both matter components are dominated by
non-relativistic matter;
\begin{align}
\rho_g=\frac{\rho_{g,0}}{a_g^3}
,\quad
\rho_f=\frac{\rho_{f,0}}{a_f^3}\,,
\end{align}
the density of dark component is approximated by
\begin{align}
\rho_{\rm D}
&=\frac{3m_f^2}{3m_f^2-2\Lambda_g}
\,\frac{K^4\rho_{f,0}}{a_f^3} \nn
&\approx 
\frac{3m_f^2}{3m_f^2-2\Lambda_g}
\,\frac{K\rho_{f,0}}{a_g^3}
+\mathcal{O}(a_g^{-6})\,.
\end{align}
Hence if $3m_f^2>2\Lambda_g$, 
$\rho_{\rm D}$ behaves as a dark matter component in the $g$-world.
If $\rho_g$ consists just of baryonic matter, 
in order to explain the observed amount of dark matter, 
we have to require
\begin{align}
\frac{\rho_{\rm D}}{\rho_g}
=\frac{3m_f^2}{3m_f^2-2\Lambda_g}
\,\frac{K\rho_{f,0}}{\rho_{g,0}}\sim 5\,.
\end{align}
With an appropriate choice of the coupling parameters, 
we find the above value, which may explain 
dark matter by the $f$-matter fluid.

\subsection{Dark matter halo} 
\label{sec_rotation_curve}
Next we discuss how 
to explain a dark matter halo around a galaxy by another one of twin matter fluids.
The existence of dark matter halo is confirmed by observations such as 
a flat rotation curve of a galaxy\cite{rotation_curve}.

Since we analyze a galactic scale, 
the background spacetime is well approximated by the 
 Minkowski metric $(\gB{}_{\mu\nu}\simeq \eta_{\mu\nu})$
ignoring the effect of a cosmological constant $\Lambda_g$.
The gravitational phenomena  can be analyzed by the linear perturbations
around the Minkowski spacetime.
 The equations of the massive mode is given by 
\begin{align}
\nablaB{}_{\mu}\nablaB{}_{\nu}h^{[-]}
-\BoxB h^{[-]}_{\mu\nu}+m_{\rm eff}^2
&\left(h^{[-]}_{\mu\nu}+\frac{1}{2}h^{[-]} \eta_{\mu\nu}
\right)
=2\Mf {}^{[-]}_{\mu\nu},\nn
\nablaB{}_{\mu} h^{[-]\mu}{}_{\nu}&=\nablaB{}_{\nu}h^{[-]}
,\nn
3m_{\text{eff}}^2h^{[-]}&=2\Mf{}^{[-]\mu}{}_{\mu}\,.
\end{align}
Substituting third equation into first one, we obtain
\begin{align}
&\nablaB{}_{\mu}\nablaB{}_{\nu}h^{[-]}
-\BoxB h^{[-]}_{\mu\nu}
+m_{\rm eff}^2h^{[-]}_{\mu\nu} \nn
&=2\Mf {}^{[-]}_{\mu\nu}-{1\over 3}\Mf{}^{[-]\rho}{}_{\rho}\eta_{\mu\nu}\,.
\label{massive_eq2}
\end{align}

To analyze the  gravitational fields of a galaxy,
we consider static Newtonian potentials $\Phi_g$ and $\Phi_f$
formed by non-relativistic mass densities $\rho_g$ and $\rho_f$.
From the $0$-$0$ component of Eq. \eqref{massive_eq2}, 
we obtain the Poisson equation for the massive mode as
\begin{align}
(\Delta -m_{\rm eff}^2 )\Phi_-=
\frac{4}{3}(4\pi G \rho_g-4\pi \mathcal{G} K^2  \rho_f) \,,
\label{massive_Poi}
\end{align}
where $\Delta=\partial^i \partial_i$ is the usual three-dimensional Laplacian operator
and $\Phi_- =-h_{00}^{[-]}/2$ is the gravitational potential of the massive mode.
The factor $4/3$ comes from van Dam-Veltmann-Zakharov (vDVZ) discontinuity 
\cite{vDVZ}.
Note that the source term is described by the difference of two mass densities,
and then it can be negative.

For the massless mode, we obtain
the ordinary form of the Poisson equation:
\begin{align}
\Delta\Phi_+
=4\pi G \frac{m_f^2}{m_{\rm eff}^2} \rho_g
+4\pi \mathcal{G} K^2\frac{m_g^2}{m_{\rm eff}^2}  \rho_f\,,
\label{massless_Poi}
\end{align}
where $\Phi_+=-h_{00}^{[+]}/2$ is the gravitational potential of the massless mode.
This source term is positive definite.

We find that both gravitational potential are affected by both 
$g$- and $f$- matter fluids.
This is main difference from the Newtonian gravity theory.
It may makes a possibility such that the $f$-matter can behave
as dark matter in the $g$-worlds.

In a small scale such as the solar system, however, GR must be restored 
because GR has been well confirmed by the experiments and observations\cite{Will}.
The restoration can be realized via the so-called Vainshtein mechanism 
\cite{original_Vainshtein}.
In this range (below the Vainshtein radius), 
the linear perturbation approach is broken down, and 
then non-linear effects must be taken into account.
However, when GR is restored from the bigravity theory,
the effect on the $g$-world  from the $f$-matter fluid  
must be screened \cite{Vainshtein}.
It indicates that the $f$-matter cannot be dark matter below the Vainshtein radius.
Since we are interested in whether the $f$-matter plays 
a role of dark matter 
in the $g$-world, we shall only analyze the linear perturbations.
The evaluation of the Vainshtein radius will be given in the last part of this subsection.

For a simplest case in which matter fluids are localized spherically, 
the Newtonian potentials are solved as
\begin{align}
\Phi_-&=\frac{4}{3}\left( \frac{G M_g}{r}e^{-m_{\rm eff}r}
-\frac{K^2\mathcal{GM}_f}{r}e^{-m_{\rm eff}r}\right)
\,,\\
\Phi_+&= \frac{m_f^2}{m_{\rm eff}^2}\frac{GM_g}{r} 
+\frac{m_g^2}{m_{\rm eff}^2} \frac{K^2\mathcal{GM}_f}{r} \,,
\end{align}
where the gravitational masses are defined by 
\begin{align}
M_g=\int 4\pi \rho_g r^2 dr
,\quad
\mathcal{M}_f=\int  4\pi \rho_f r^2 dr\,.
\end{align}
From \eqref{g_f_perturbation}, 
the Newtonian potentials in the  $g$- and $f$-worlds
are described as
\begin{align}
\Phi_g&=\Phi_++{m_g^2\over m_{\rm eff}^2}\Phi_-
\nn
&=-\frac{GM_g}{r} \left(
\frac{m_f^2}{m_{\rm eff}^2}
+\frac{4m_g^2}{3m_{\rm eff}^2}e^{-m_{\rm eff}r} \right) 
\nn
&~~~~-\frac{m_g^2}{m_{\rm eff}^2}\frac{K^2\mathcal{GM}_f}{r}
\left(1-\frac{4}{3}e^{-m_{\rm eff}r}\right)
\label{g_potential}\,,
\\
\Phi_f&=\Phi_+-{m_f^2\over m_{\rm eff}^2}\Phi_-
\nn
&=-\frac{K^2\mathcal{GM}_f}{r} \left(
\frac{m_g^2}{m_{\rm eff}^2}
+\frac{4m_f^2}{3m_{\rm eff}^2}e^{-m_{\rm eff}r} \right)
\nn
&~~~~-\frac{m_f^2}{m_{\rm eff}^2}\frac{GM_g}{r}
\left(1-\frac{4}{3}e^{-m_{\rm eff}r}\right)\,.
\end{align}
where $\Phi_g =-h_{00}^{[g]}/2,\Phi_f=-h_{00}^{[f]}/2$.

Let us consider the Newtonian potential in the $g$-world.
Below the Compton wavelength of the massive graviton
($r<m_{\rm eff}^{-1}$), 
the potential becomes
\begin{align}
\Phi_g&=-\frac{GM_g}{r} \left(
1
+\frac{m_g^2}{3m_{\rm eff}^2} \right)
+\frac{m_g^2}{3m_{\rm eff}^2}\frac{K^2\mathcal{GM}_f}{r}\,.
\end{align}
Note that the second term is positive definite.  It means
that the $f$-matter acts as a repulsive force in the $g$-world.
It comes from the factor $4/3$ in \eqref{g_potential}.
To explain dark matter, of course, the gravitational force must be attractive.
Therefore, the $f$-matter cannot behaves as 
dark matter when the size of the localized system is smaller than
 the Compton wavelength.

The origin of this repulsive force is the massive mode, which
cannot propagate in the large system such that $m_{\rm eff}r \gg 1$.
In fact, beyond the  Compton wavelength ($r>m_{\rm eff}^{-1}$), 
the potential is approximated by
\begin{align}
\Phi_g=-\frac{G_{\rm eff}}{r}(M_g+K^4\mathcal{M}_f)
\label{Newton_potential}
\end{align}
where
\begin{align}
G_{\rm eff}=\frac{m_f^2}{m_{\rm eff}^2}G
\end{align}
is the local effective gravitational constant.
This potential is formed by the $f$-matter 
as well as the $g$-matter.
Hence, it is possible to explain dark matter 
by another one of twin matter fluids.

Inside the Vainshtein radius,  the gravitational constant is restored to 
the Newtonian gravitational constant \cite{Vainshtein}.
Since the difference between the effective gravitational constant  at a galactic scale
and the Newtonian one should not be so large \cite{Will}, 
we find a constraint such that
\begin{align}
\frac{m_g^2}{m_f^2}=\frac{K^2\kappa_g^2}{\kappa_f^2} \ll 1\,.
\label{kappa_constraint}
\end{align}

\begin{figure}[tbp]
  \centering
  \includegraphics[width=6.5cm]{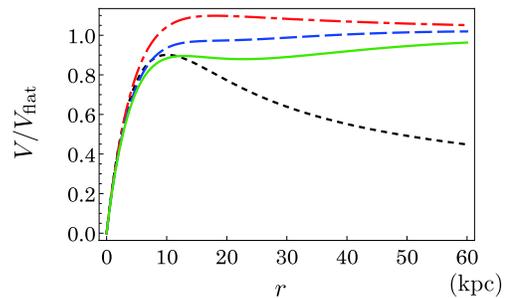}
  \caption{The rotation curve in the $g$-worlds. 
We plot three cases of $m_{\rm eff}^{-1}=5$ (the red dashed-dotted curve), 
$10$ (the blue dashed curve) and $15$ kpc (the green solid curve). 
Matter distributions are given by $\rho_g(r)=\rho_{g}(0)\exp[-r/r_{\rm gal}],
\rho_f(r)=\rho_{f}(0)(1+(r/r_{\rm halo})^2)^{-1}$, 
where we set $r_{\rm gal}=r_{\rm halo}=3$ kpc and $\rho_g(0)=\rho_f(0)$.
The effective gravitational constant is $G_{\rm eff}/G=0.961538$ ($m_g/m_f=0.2$).
The black dotted curve is the rotation curve without $f$-matter.}
\label{rotation_curve}
\end{figure}

Now we check whether the rotation curve becomes flat at a galaxy scale or not.
For simplicity, we assume a spherically symmetric matter distribution as 
\begin{align}
\rho_g(r)&=\rho_{g}(0)\exp[-r/r_{\rm gal}],\nn
\rho_f(r)&=\frac{\rho_{f}(0)}{1+(r/r_{\rm halo})^2}\,.
\label{matter_dist}
\end{align}
We show the resulting rotation curves for several values of $m_{\rm eff}$
 in   Fig. \ref{rotation_curve}.
The rotation velocity $V$ is evaluated as $V^2=rd\Phi_g/dr$. 
We find a flat rotation curve if $m_{\rm eff}^{-1} \sim $ kpc.
Note that since
\bea
m_{\rm eff}^{-1} =6.4\times \left(
{m_{\rm eff}\over 10^{-33}{\rm eV}}
\right)^{-1}\,{\rm Gpc}
\,,
\ena
we have the solid limit on the Compton wave length as
\bea
m_{\rm eff}^{-1} > 0.091 \,{\rm pc}
\ena
from the experimental constraint on the graviton mass \cite{graviton-mass, Will}.

We then conclude that
the $f$-matter behaves as dark matter  in the $g$-world
 if the Compton wave length of the massive graviton is less than 
a galaxy scale such as $m_{\rm eff}^{-1}\sim 1$ kpc.
When the mass becomes lighter, then the rotation velocity decreases.
It is due to a ``repulsive force" induced by the massive mode 
because the Compton wavelength becomes larger.
Note that in the shorter range than $r\sim 10$, the rotational velocity 
with the $f$-matter (the green curve) 
is smaller than that without the $f$-matter (the black dotted curve),
which is the evidence that the $f$-matter acts as a repulsive force.

In order to justify the above analysis, we have to 
 evaluate the Vainshtein radius below which the linear approximation is
broken.
Performing  the same method as \cite{Vainshtein}, 
we find the linear perturbation analysis
for a spherically symmetric system is valid only when
\begin{align}
m_{\rm eff}^2 &\gg \frac{GM_-(r)}{r^3} \,, 
\end{align}
where
\begin{align}
GM_-(r)&:=  \left|
G\int^r_0 4\pi \tilde{r}^2 \rho_g(\tilde{r}) d\tilde{r}
-
K^2\mathcal{G}\int^r_0 4\pi \tilde{r}^2 \rho_f(\tilde{r})  d\tilde{r}
\right|\,.
\end{align}
Here we have ignored a cosmological constant.
The mass of galaxy is dominated by the dark matter component, and 
 we have the constraint \eqref{kappa_constraint},
we find $K^2\mathcal{GM}_f \gg GM_g$,
 where $M_g$ and $\mathcal{M}_f$ 
are total masses of the $g$- and $f$-matter fluids, respectively.
Hence the right hand side is bounded from the above as 
\begin{align*}
&GM_-(r)
\leq K^2\mathcal{GM}_f
\,.
\end{align*}

As a result, we conclude that
the linear perturbation analysis is valid for 
\begin{align}
r &\gg r_{V}:=\left( \frac{K^2\mathcal{GM}_f}{m_{\rm eff}^2}\right)^{1/3} \nn
 \,.
\end{align}
From Eq. (\ref{Newton_potential}), we find  the effective mass of a galaxy
 in the $g$-world is 
\bea
M_{\rm gal}\approx {m_f^2\over m_{\rm eff}^2}
K^4\mathcal{M}_f \,.
\ena
For $M_{\rm gal}\sim 10^{12} M_{\odot}$, we can evaluate the Vainshtein radius as
\begin{align}
r_{V}&\sim 0.04 \; {\rm kpc} \;  
\left( \frac{m_{\rm eff}^{-1}}{1 \; {\rm kpc}} \right)^{2/3}
 \left( \frac{1}{1-G_{\rm eff}/G} \right)^{1/3}
 \,.
\end{align}
It guarantees that the linear perturbation approximation is valid in a galactic scale
if $m_{\rm eff}^{-1} \lesssim$ kpc. 

Such a galactic scale graviton mass 
as well as a cosmological constant to explain dark energy
can be obtained if the ratio of two 
gravitational constants is given by $\kappa_f^2/\kappa_g^2 \sim 10^{12}$
as shown in Appendix \ref{eff_mass_cc}.
However, 
the linear perturbation approximation may not be 
justified because 
\bea
r_V\sim 0.4 {\rm Mpc} \times \left( \frac{m_{\rm eff}^{-1}}{1 \; {\rm kpc}} \right)^{2/3}
 K^{-3/2}
\,,
\ena
which may give the larger Vainshtein radius such as 1 Mpc unless 
 $K\gg O(1)$.
We may have to fine-tune the coupling constants $\{b_i\}$ as shown in 
 Appendix \ref{eff_mass_cc}.

\subsection{Cosmic structure formation}
\label{sec_structure_formation}

Finally, we discuss
the evolution of cosmological density perturbations
based on the linear perturbation theory \cite{cp_theory}.
For simplicity, we assume that the background  flat FLRW spacetimes are given by
the homothetic solutions.
We shortly summarize the perturbation equations 
in Appendix \ref{comological_perturbation}.

\subsubsection{Numerical solutions}

Since we are interested in formation of galaxies, 
we discuss only sub-horizon scale perturbations, $a/k \ll H^{-1}$. 
In this subsection, we first analyze the linear perturbation equations numerically.
We assume that  the matter component  is dominated by non-relativistic matter  $(w=0)$.
Since there is another scale of length, i.e., the Compton wave length of the massive 
graviton $m_{\rm eff}^{-1}$,
we can classify those three scales into three possibilities: \\[.5em]
\indent Case (a) $a/k \ll H^{-1} \ll m_{\rm eff}^{-1}$, \\[.5em]
\indent Case (b) $a/k \ll m_{\rm eff}^{-1} \ll H^{-1}$, \\[.5em]
\indent Case (c) $m_{\rm eff}^{-1} \ll a/k \ll H^{-1}$.\\[.5em]
~~
Assuming the initial data at the decoupling time is given in each case,
we solve  the perturbation equations \eqref{cp_eq1}-\eqref{cp_eq7} numerically.

\begin{figure}[h]
\begin{center}
  \includegraphics[width=5.5cm]{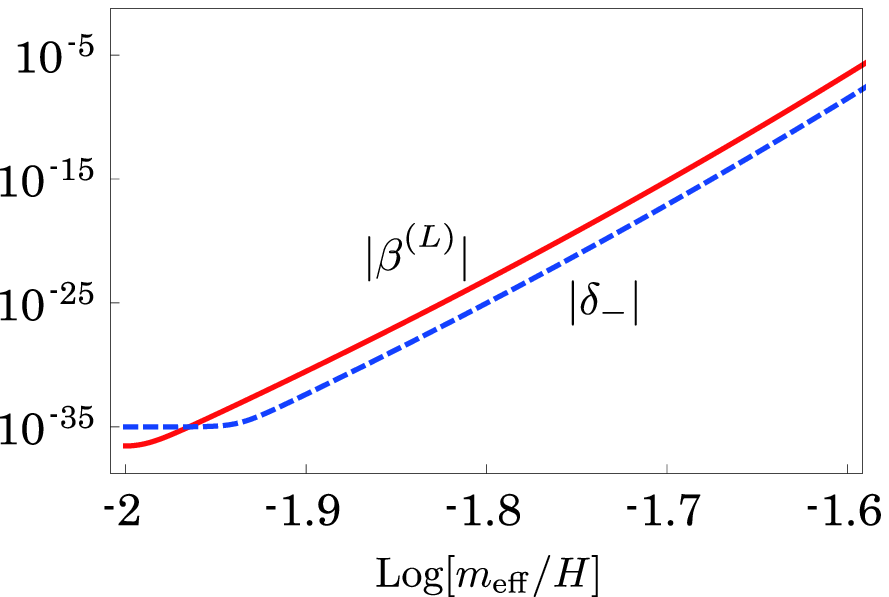}
  \\
  (a) $a_{\rm in}/k \ll H_{\rm in}^{-1} \ll m_{\rm eff}^{-1}$
  \\
  \vspace{3mm}
  \includegraphics[width=5.5cm]{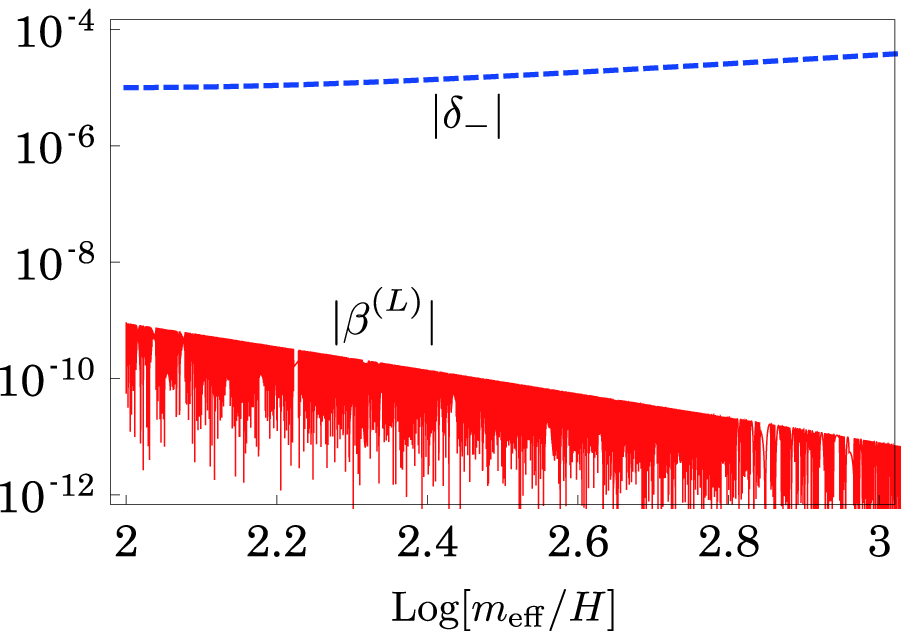}
  \\
  (b) $a_{\rm in}/k \ll m_{\rm eff}^{-1} \ll H_{\rm in}^{-1}$ 
  \\
  \vspace{3mm}
  \includegraphics[width=5.5cm]{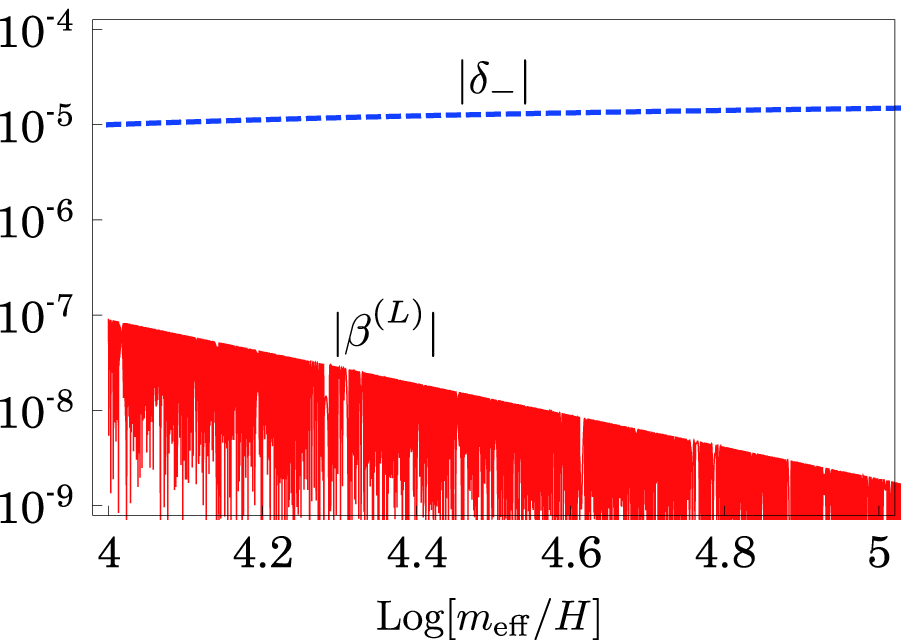}
  \\
  (c) $m_{\rm eff}^{-1} \ll a_{\rm in}/k \ll H_{\rm in}^{-1}$
  \\
  \caption{The time evolution of $\beta_-^{(L)}$ 
and $\delta_-$.
The background spacetime is the dust dominant universe $(a\propto t^{2/3})$.
We choose 
the initial data 
(a) $a_{\rm in}/k=10^{-4}\times m_{\rm eff}^{-1}, H_{\rm in}^{-1}=10^{-2}\times m_{\rm eff}^{-1}$, 
(b) $a_{\rm in}/k=10^{-2}\times m_{\rm eff}^{-1}, H_{\rm in}^{-1}=10^{2}\times m_{\rm eff}^{-1}$, and 
(c) $a_{\rm in}/k=10^{2}\times m_{\rm eff}^{-1}, H_{\rm in}^{-1}=10^{4}\times m_{\rm eff}^{-1}$.
The perturbations grow exponentially for (a). 
For (b) and (c), the metric perturbation $\beta_-^{(L)}$ decays
with oscillations, while 
the density perturbation $\delta_-$ 
increases slowly without oscillation.}
  \label{cp_fig}
\end{center}
\end{figure}

We show the results in Fig. \ref{cp_fig},
 where we have chosen 
the initial data as 
(a) $a_{\rm in}/k=10^{-4}\times m_{\rm eff}^{-1}, H_{\rm in}^{-1}=10^{-2}\times m_{\rm eff}^{-1}$, 
(b) $a_{\rm in}/k=10^{-2}\times m_{\rm eff}^{-1}, H_{\rm in}^{-1}=10^{2}\times m_{\rm eff}^{-1}$, and 
(c) $a_{\rm in}/k=10^{2}\times m_{\rm eff}^{-1}, H_{\rm in}^{-1}=10^{4}\times m_{\rm eff}^{-1}$.
We show two variables; one metric component $\beta^{(L)}$ and 
the density perturbation $\delta_-$.
In the calculation, we have ignored the terms with the sound speed 
because we consider the perturbations  larger  than
the Jeans length, 
i.e. $k \ll k_J =a \sqrt{ 4\pi G \bar{\rho}}/c_s$.

For the case (a), both perturbation variables ($\beta^{(L)}, \delta_-$) grow exponentially.
Hence the linear perturbation  is unstable. 
On the other hand, for the cases (b) and (c), 
the metric perturbation $\beta_-^{(L)}$ decays
with oscillations, which frequency is about $\sqrt{(k/a)^2+m_{\rm eff}^2}$, while 
the density perturbation $\delta_-$ 
increases monotonically without oscillations.
The increase rates are evaluated numerically by power-law functions of the scale 
factor $a$ as 
$\delta_-\propto a^{1.176}$ and $ a^{0.1077}$ for (b) and (c),
respectively.
 
The Compton wavelength  $m_{\rm eff}^{-1}$ 
is larger than the horizon scale $H^{-1}$ for (a),
while the relation is opposite for (b) and (c).
Hence the above result concludes that if  $m_{\rm eff}^{-1}>H^{-1}$ (the case (a)),
the perturbative approach is no longer valid.
Note that it was shown that in the bigravity theory 
there exists a gradient instability against linear cosmological perturbations
 in the massless limit \cite{cp_instability1,cp_instability2,cp_instability3}.
The non-linear effect must be taken into account.

When  $m_{\rm eff}^{-1}<H^{-1}$ (the case (b) and (c)), 
 there are two important time scales: One is the Hubble expansion time 
$H^{-1}$, and the other is the oscillation time scale of the massive 
graviton $m_{\rm eff}^{-1}$.
We find that the metric variables $\{ \alpha_{-}, \beta_{-}^{(L)}, h_{-}^{(L)}, h_{-}^{(T)}\}$
 are divided into two parts; 
 the monotonically growing part and 
the oscillating part. 
The former part changes in the Hubble expansion time $H^{-1}$, 
while the latter part with the high frequency  $\sqrt{(k/a)^2+m_{\rm eff}^2}$
 is always decaying. 
The metric component $\beta_-^{(L)}$ has no former part, and then 
eventually vanishes as shown in In Fig. \ref{cp_fig}. 
On the other hand,  the matter perturbations $\{\delta_{-}, v^{(L)}_{-}\}$
grow slowly in the Hubble time scale $H^{-1}$  without oscillation.

As a result, all variables 
asymptotically approach monotonic functions increasing
in  the Hubble time scale $H^{-1}$.
There seems to exist an asymptotic solution which changes 
monotonically in  the Hubble time scale $H^{-1}$.
We then assume that the perturbation variables change 
in the Hubble time scale $H^{-1}$, i.e., 
 $|\dot{X}_-| \sim |H X_-|$, which provides 
the above asymptotic solution.
We call such an approach an adiabatic potential approximation
\cite{footnote2}, 
since we ignore the oscillation parts of metric which correspond to 
the scalar gravitational waves.

\subsubsection{Adiabatic potential approximation}
Under the adiabatic potential approximation,
we look for a solution for sub-horizon scale perturbations.
From the perturbation equations for 
the massive mode, \eqref{cp_eq1}, \eqref{cp_eq6} and \eqref{cp_eq7},
we find
\begin{align}
-\left(2\frac{k^2}{a^2}+3m_{\rm eff}^2 \right)
\alpha_-
&=\kappa_g^2\bar{\rho}_g \delta_- +3m_{\rm eff}^2 h_-^{(L)},
\label{massive_P1}
\\
\beta_-^{(L)}=0\,, \quad
h_-^{(T)}&=-3\left(\frac{\alpha_-}{2} +h_-^{(L)}\right)\,.
\end{align}
Substituting \eqref{cp_trace_eq} into \eqref{massive_P1},
we obtain
\begin{align}
-\left( \frac{k^2}{a^2}+m_{\rm eff}^2\right) \alpha_-
=\frac{4}{3}\times
 \frac{\kappa_g^2\bar{\rho}_g}{2} \delta_-\,,
\label{massive_P3}
\end{align}
where we have ignored a cosmological constant compared with the graviton mass term,
because we are interested in the case with a rather large value of  $m_{\rm eff}$.
This equation is interpreted as the massive Poisson equation.
The factor $4/3$ comes from the vDVZ discontinuity.
Using Eq. (\ref{massive_P3}) and ignoring the sound velocity term, 
the equation for the density perturbation $\delta_-$ is described as
\begin{align}
\ddot{\delta}_- +2 H\dot{\delta}_- 
-\frac{4k^2/a^2}{3(k^2/a^2+m_{\rm eff}^2)} 
\frac{\kappa_g^2\bar{\rho}_g}{2}
\delta_-=0
\label{massive_eom_delta}\,.
\end{align}

As we showed numerically, the solution of this equation is 
found as an attractor for generic initial data
 if $m_{\rm eff}^{-1}<H^{-1}$ is satisfied initially.
However, the condition
$m_{\rm eff}^{-1}<H^{-1}$  is not always true.
In fact, when we go back to the past, since $H^{-1}\sim t$, 
then  the condition is broken in the past epoch.

When we start from the epoch of $m_{\rm eff}^{-1}>H^{-1}$,
which corresponds to the case (a), 
the linear perturbation is unstable, and then 
non-linear effect must be taken into account. 
We can see this fact from the constraint equation.
For the small scale such that $a/k \ll m_{\rm eff}^{-1}(<H^{-1})$,
 we find 
\begin{align}
\left[3m_{\rm eff}^2-2\Lambda_g-(1-w)\kappa_g^2\bar{\rho}_g \right]h_-^{(L)}
=\frac{\kappa_g^2\bar{\rho}_g}{3}(\delta_- -3w\pi^{(L)}_-)\,.
\label{cp_eq8}
\end{align}
 from \eqref{cp_trace_eq}. 
Note that $(3m_{\rm eff}^2-2\Lambda_g)$ is a positive constant
if the Higuchi bound is satisfied, 
while $-(1-w)\kappa_g^2\bar{\rho}_g$ 
for the ordinary matter is negative definite and its magnitude 
decreases  in time.
Hence the coefficient of the left hand side of \eqref{cp_eq8} 
eventually vanishes when we go back to the past,
while the right hand side does not usually vanish simultaneously.
It indicates that the linear perturbation approximation is broken at the 
time when the coefficient of the left hand side vanishes
because $h^{(L)}_-$ must diverge.
In this epoch, to answer for the question 
whether there still exists an adiabatic potential  solution 
as an attractor, 
we have to analyze the full dynamical equations with inhomogeneities,
which is quite difficult without heavy numerical simulation.
However, there is some hope from Eq. \eqref{cp_eq8},
which shows a possibility such that  
the density perturbation is still small enough to 
 be treated as linear perturbation
even when the metric perturbations become nonlinear.
In a spherically symmetric case,
 we find an adiabatic potential  solution with nonlinear metric perturbations 
but with  linear matter perturbations\cite{AM3}.
In this solution, we claim that  the Vainshtein mechanism 
is working even in a cosmological context,
and the solution can be described  by GR.

\begin{figure}[htbp]
\begin{center}
  \includegraphics[width=8cm]{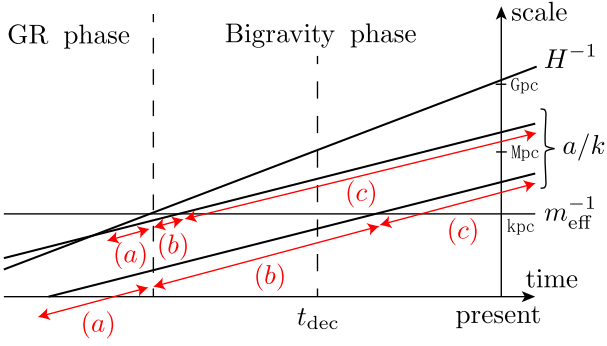}
  \caption{The schematic diagram of the growth history.
In the early stage of the Universe ($H^{-1}<m_{\rm eff}^{-1}$),
because of the Vainshtein mechanism,
the standard big bang universe is recovered.
However the Universe eventually evolves into the bigravity phase,
 in which there are two cases (b) and (c) depending on 
the perturbation scale compared with $m_{\rm eff}^{-1}$.}
  \label{scenario}
\end{center}
\end{figure}

Hence we may conceive the following scenario, although 
the present analysis is based on the perturbations around a homothetic solution 
and an extended analysis with more generic background such as that in 
\cite{cp_instability2} will be required.
In the early stage of the universe, because of the Vainshtein mechanism, 
gravity is described by GR and then the standard big bang scenario is found. 
However the Universe eventually evolves into the bigravity phase at 
$H^{-1}\sim m_{\rm eff}^{-1}$ as
shown in Fig. \ref{scenario}.
When the universe reaches the decoupling time, 
we find the case (b) or (c) for the perturbations, in which 
the adiabatic potential approximation becomes valid as an attractor.
Hence we analyze whether the $f$-matter can be 
dark matter in  the cosmic structure formation,
using the above  adiabatic potential approximation.

\subsubsection{Growth history of density perturbation}
The evolution equation of density perturbation
for the massless mode in a sub-horizon scale 
is given from Eq. (\ref{density_perturbation_massless}) as 
\begin{align}
\ddot{\delta}_+ +2H \dot{\delta}_+ -\frac{\kappa_g^2\bar{\rho}_g}{2} \delta_+&=0 
\label{massless_eom_delta}
\end{align}
where we have ignored a cosmological constant and the term with a sound velocity
 as before.
From Eqs. (\ref{massive_eom_delta}) and (\ref{massless_eom_delta}), 
we obtain the equations for the physical density perturbations
 ($\delta_g$ and $\delta_f$)  as
\begin{align}
&\ddot{\delta}_g +2H \dot{\delta}_g -4\pi G_{\rm eff}
(\bar{\rho}_g\delta_g
+\bar{\rho}_{\rm D}\delta_{\rm D}
)=0\,,
\nn
&\ddot{\delta}_f+2H\dot{\delta}_f-4\pi \mathcal{G}_{\rm eff}
(\bar{\rho}_f\delta_f
+\bar{\rho}_{\rm G}\delta_{\rm G}
)=0\,,
\end{align}
where
\begin{align}
G_{\rm eff}&=G \frac{m_f^2}{m_{\rm eff}^2}
\left(1
+\frac{m_g^2}{m_f^2} F\right)
\,,\\
\bar{\rho}_{\rm D}&=K^4\bar{\rho}_f
\,,\\
\delta_{\rm D}&=\frac{1-F }
{1+{m_g^2\over m_f^2}F }\delta_f
\,, 
\end{align}
and
\begin{align}
\mathcal{G}_{\rm eff}&=\mathcal{G} \frac{K^2m_g^2}{m_{\rm eff}^2}
\left(1
+\frac{m_f^2}{m_g^2} F\right)
\,,\\
\bar{\rho}_{\rm G}&=K^{-4}\bar{\rho}_g
\,,\\
\delta_{\rm G}&=\frac{1-F }
{1+{m_f^2\over m_g^2}F }\delta_g\,,
\end{align}
with 
\begin{align}
F:=\frac{4m_{\rm eff}^{-2}}{3(m_{\rm eff}^{-2}+a^2/k^2)} 
\,.
\end{align}

Beyond the  Compton wavelength of the massive graviton,
the effective gravitational constant becomes $G_{\rm eff}/G
\approx m_f^2/m_{\rm eff}^2$.
It is the same not only as the cosmological value but also
as the local one if the  graviton mass is large ($m_{\rm eff}^2
\gg \Lambda_g$).
The perturbation of dark matter component coincides with 
that of the $f$-matter, i.e.,
\begin{align}
\delta_{\rm D} \approx \delta_f\,,
\end{align}
for $a/k \gg m_{\rm eff}^{-1}$.
Therefore, the $f$-matter perturbation 
behaves as the dark matter component in the $g$-world
as \S. \ref{sec_cosmic_pie} and \S. \ref{sec_rotation_curve}.

Inside the Compton wavelength, the $f$-matter acts as a repulsive force
as shown in  \S. \ref{sec_rotation_curve}.  
In the present case, it can be seen explicitly from the relation 
\bea
\delta_{\rm D}\sim -{1\over 3+4{m_g^2\over m_f^2}}\delta_f
\ena
 for $a/k \ll m_{\rm eff}^{-1}$.
It indicates that the $g$-matter  accumulates in a  low-density region of 
the $f$-matter.

\begin{figure}[htbp]
\begin{center}
  \includegraphics[width=5.5cm]{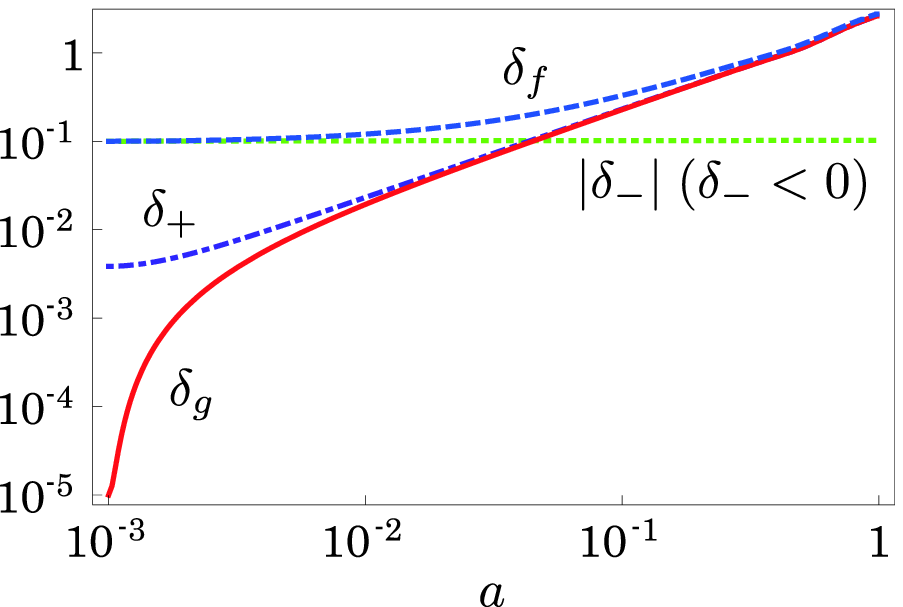}
  \\
  (a) $k^{-1}=10\, {\rm Mpc}$
  \\
  \vspace{3mm}
  \includegraphics[width=5.7cm]{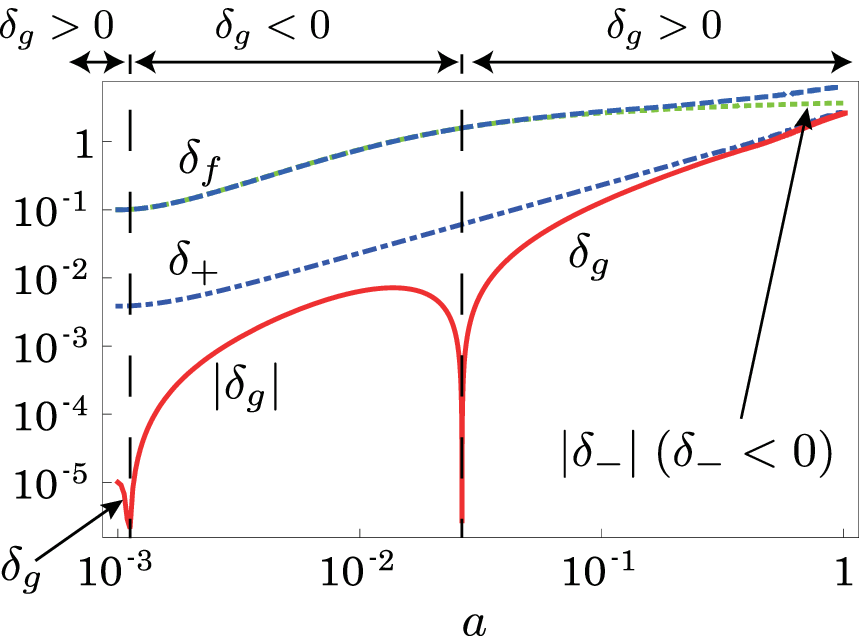}
  \\
  (b) $k^{-1}=10^2\, {\rm kpc}$
  \\
  \caption{The evolution of density perturbations for two scales
[(a) 10Mpc and (b) 100kpc at the present ($a=1$)]. 
We assume $\delta_g=10^{-5}$ and 
$\delta_f=10^{-1}$ at the decoupling time ($a=10^{-3}$).
 The blue dashed curve indicates the evolution of $\delta_f$,
while the red solid curve indicates that of $\delta_g$.
We set $m_{\rm eff}^{-1}=1\, {\rm kpc}$ and $m_g/m_f=0.2$.
The background spacetime is the dust dominant universe $(a\propto t^{2/3})$.
 }
\label{linear_cp}
\end{center}
\end{figure}

We show the numerical result of 
the evolution of density perturbations for two different scales
[$k^{-1}=$ 10Mpc and 100kpc at the present ($a=1$)]
in Fig. \ref{linear_cp}.
We assume $\delta_g=10^{-5}$ and 
$\delta_f=10^{-1}$ at the decoupling time ($a=10^{-3}$).
For the large scale perturbation, its scale is
 always larger than the Compton wavelength after the decoupling time.
Hence the $f$-matter plays the role of dark matter in the $g$-world
and helps small baryon perturbation $\delta_g$ to grow rapidly as
 shown in Fig. \ref{linear_cp} (a).
The evolution of $\delta_g$ 
is similar to the growth of density perturbations with CDM in GR.

On the other hand, for the small scale perturbation,
its scale is shorter than the Compton wavelength at the decoupling time.
During the period of $a/k<m_{\rm eff}^{-1}$,
the $f$-matter acts as a repulsive source in the $g$-world.
Then the evolution of $\delta_g$ is quite different  due to the appearance of 
a repulsive force by the $f$-matter perturbations
as shown in Fig. \ref{linear_cp} (b).
 $\delta_g$  changes its  sign and then 
decreases to a negative value in the early stage.
But the perturbation scale eventually exceeds $m_{\rm eff}^{-1}$
 as the scale factor increases.
In fact the perturbation scale becomes larger than the Compton wavelength 
after  $a=k/\sqrt{3} \times m_{\rm eff}^{-1}$, 
when $\delta_{\rm D}$ changes its sign.
After then, 
the $f$-matter begins to act as dark matter.
As shown in Fig. \ref{linear_cp} (b),
 $\delta_g$  changes its sign again to be positive.
$\delta_g$ then grow into a nonlinear regime
via large density perturbations of the $f$-matter fluid.

We set $m_g/m_f=0.2$, which satisfies  the constraint \eqref{kappa_constraint}.
From Eq. (\ref{gf+-}), we find that 
the perturbations of the $g$-variables are dominated by the massless mode, 
while those of the $f$-variables have a significant influence by the massive mode.
Since the massive mode can grow only when $a/k \ll m_{\rm eff}^{-1}$,
$\delta_f$ grows  first and then 
$\delta_g$ follows as shown in Fig. \ref{linear_cp}  (b).
On the other hand,  as shown in Fig. \ref{linear_cp} (a),
$\delta_f$ cannot grow at first
because the massive mode cannot grow for $a/k \gg m_{\rm eff}^{-1}$.
$\delta_f$ starts to grow
after the perturbation of the massless mode catches up to that of the massive mode.
$\delta_g$ grows rapidly by the increase of the massless mode even when $\delta_f$ 
does not grow.

We  conclude that the cosmic structure formation can  also be explained by 
another one of twin matter fluids.

\section{Concluding Remarks}
\label{summary}
We have studied a possibility to explain a dark matter component 
by another one of twin matter fluids
in the ghost free bigravity theory.
We have analyze from a galactic scale to a cosmological scale.
If we assume the Compton wavelength of the massive graviton is
shorter than a galactic scale, 
i.e.,  a  graviton mass is rather heavy 
($m_{\rm eff}\sim 10^{-27}$ eV),
we find  a dark matter component can be explained by another twin matter 
for all scales.
For such a model, our matter field consists just of baryons.
The origin of dark matter field is another matter field 
which couples only to another metric.

For such a model, at first glance, it seems that 
the cosmic acceleration cannot be
explained by the interaction term,
because the expected cosmological constant is also large.
As shown in Appendix A (see Table \ref{heavy}), however, 
we can  find a large graviton mass and a small cosmological constant
although we need a fine-tuning of the coupling parameters. 
For such fine-tuned coupling parameters, the ghost-free bigravity theory could
explain dark matter as well as dark energy. 

Our analysis is valid only for the late stage of the Universe, because 
the background space is assumed to be homothetic. 
In order to find a whole history of the Universe, 
we have to analyze more generic background spacetime with 
perturbations.  We also have to show that the Vainshtein mechanism 
does really work in the early stage of the Universe as we conceived.
Those are in progress. 

Our result shows the graviton mass is phenomenologically significant.
The bigravity theory can explain only dark energy for
$m_{\rm eff}^{-1} \sim$ Gpc,
while if $m_{\rm eff}^{-1} \lesssim$ kpc
it could explain dark matter (as well as dark energy).
Therefore, an important remaining question is 
how large graviton mass is possible from 
the theoretical and observational points of view. 
From the theoretical point of view, 
we should start from more fundamental theory 
in which a bigravity theory is reduced
as a low-energy effective theory \cite{h_dim1,h_dim2,h_dim3}.
We hope that
the hierarchy between the graviton mass and an 
effective cosmological constant to explain both dark sectors 
will be solved in such a fundamental theory.
From the observational point of view, 
the evidence of a graviton mass could be detected by gravitational waves 
\cite{Will, GW_bigravity}.
Furthermore, comparing some bigravity phenomena with the observational data
in a galactic scale as well as in a cosmological scale, we may find
 the constraint of the graviton mass 
(e.g. gravitational lensing by galaxies \cite{lensing}).
In order to clarify whether the graviton really has a mass 
and give a constraint on its value,
further studies are required.

\section*{Acknowledgments}

We would like to thank Hideo Kodama, Norihiro Tanahashi and Ayumu Terukina for useful discussions.
KM would like to thank DAMTP, the Centre for Theoretical Cosmology,
and Clare Hall in the University of Cambridge, where this work was completed.
This work was supported in part by Grants-in-Aid from the 
Scientific Research Fund of the Japan Society for the Promotion of Science 
(No. 25400276). 

\newpage

\newpage

\appendix
\section{Evaluation of the effective cosmological constant and the graviton mass}
\label{eff_mass_cc}
The effective cosmological constant and the graviton mass 
are given by \eqref{eff_cc} and \eqref{eff_mass}, 
which contain many unknown or unfixed values of coupling constants. 
In order to discuss the evolution of the Universe, 
we have first to evaluate the values of the
graviton mass and the cosmological constant
for given coupling constants.

For this purpose, it is more convenient to introduce another set of 
coupling constants $\{c_k \} (k=0,1, \cdots, 4)$
 by rewriting the interaction term in term of another tensor defined by
$\mathcal{K}^{\mu}_{\nu}=\delta^{\mu}_{\nu}-{\gamma^{\mu}}_{\nu}$
as 
\begin{equation}
\mathscr{U}(g,f)=\sum^4_{k=0}c_k\mathscr{U}_k(\mathcal{K})
\,.
\end{equation}
The relations between $\{b_k\}$ and $\{c_k\}$ are given by
{\setlength\arraycolsep{2pt}\begin{eqnarray}
c_0&=&b_0+4b_1+6b_2+4b_3+b_4, \nonumber \\
c_1&=&-(b_1+3b_2+3b_3+b_4), \nonumber \\
c_2&=&b_2+2b_3+b_4, \\
c_3&=&-(b_3+b_4), \nonumber \\
c_4&=&b_4. \nonumber
\end{eqnarray}}
We assume that a flat Minkowski spacetime exists in the 
present bigravity model. Then 
we impose the following conditions:
\begin{equation}
c_0=c_1=0
\,.
\label{c_0=0}
\end{equation}
If $m$ is assumed to be 
the graviton mass  in the Minkowski background 
 in massive gravity limit, 
we should set  
\begin{equation}
c_2=-1
\,.\label{c_2=-1}
\end{equation}
As a result, $\{b_k\}$ are described by 
two free coupling constants $c_3$ and $c_4$ as
{\setlength\arraycolsep{2pt}\begin{eqnarray}
b_0&=&4c_3+c_4-6, \nonumber \\
b_1&=&3-3c_3-c_4, \nonumber \\
b_2&=&2c_3+c_4-1, \label{coupling_Minkowski}\\
b_3&=&-(c_3+c_4), \nonumber \\
b_4&=&c_4\,.
\nonumber 
\end{eqnarray}}
These coupling constants guarantee
\begin{align}
m_{\rm eff}^2|_{K=1}=m^2,
\quad
\Lambda_g|_{K=1}=\Lambda_f|_{K=1}=0\,,
\end{align}
for the Minkowski background with $K=1$.

In order to explain dark energy, 
de Sitter spacetime must be an attractor solution.
As shown in \cite{with_twin_matter, homothetic}, 
the quartic equation (\ref{eq_K}) gives one de Sitter solution with $K=K_{\rm dS}$ as 
well as two anti de Sitter solutions, if 
\begin{align}
2c_3^2+3c_4>0\,.
\end{align}
Since the Higuchi bound  must be satisfied\cite{Higuchi}, 
the lower bound of the graviton mass is given by the cosmological constant as
\begin{align*}
m_{\rm eff}^2 > \frac{2}{3}\Lambda_g \,.
\end{align*}

If we consider a simple and natural case, 
i.e., $b_k\sim O(1)$ (or $c_k\sim O(1)$) and $\kappa_g\sim\kappa_f$,
we find the cosmological constant and the graviton mass as
\begin{align}
\Lambda_g &\sim m^2
\nn
m_{\rm eff} &\sim m
\end{align}
for $K=K_{\rm dS}$, assuming no fine-tuning of the coupling constants.

In this case, dark energy fixes the value of $\Lambda_g$,
and then $m^{-1}$ (the Compton wave length)
must be the cosmological horizon scale $H^{-1}$. 
As a result, the massive mode becomes important for a sub-horizon scale 
such as a galaxy. 
In this case, the $f$-matter does not explain dark matter, because 
it is in the GR phase. 
In order for the $f$-matter to be dark matter, 
$m_{\rm eff}\sim m$ is too light. 
As we show in \S. \ref{origin_dark_matter},
if $m_{\rm eff}\sim $ 1 kpc,  the $f$-matter 
can play a role of dark matter. However, in that case, 
$\Lambda_g$ is too large to explain the cosmic acceleration,
except for the $K=0$ branch with a different origin of dark energy.

Is there any possibility such that 
$\Lambda_g\sim H^{-1}$ but $m_{\rm eff}\sim $ 1 kpc ?
We then look for the possibility of a heavy graviton mass, 
i.e. $\Lambda_g \ll m_{\rm eff}^2$.
One way to get a heavy graviton mass as well as a small cosmological constant
is to assume $\kappa_g^2 \gg \kappa_f^2$ or $\kappa_f^2 \gg \kappa_g^2$.
If we have such a hierarchy between two gravitational constants,
 we find $\Lambda_g \ll m_{\rm eff}^2$ without fine-tuning of 
coupling constants $\{c_i\}$.
Otherwise, we have to fine-tune the coupling constants.
Fine-tuning the coupling constants such that 
\begin{align*}
0<2c_3^2+3c_4 \ll 1\,,
\end{align*}
we find 
a small effective cosmological constant 
$(\Lambda_g \ll m_{\rm eff}^2\sim m^2)$.
We show some examples  in Table \ref{heavy}.

\begin{table}[h]
\begin{center}
    \caption{The ratios of the  cosmological constant to the graviton mass square. 
We assume $c_3=-1$.}
    \tabcolsep =2mm
    \renewcommand{\arraystretch}{1.3}
  \begin{tabular}{cccccccccc}
\hline
$\kappa_g^2/\kappa_f^2$ & $2c_3^2+3c_4$ & $K_{\rm dS}$ &
$\Lambda_g/m_{\rm eff}^2$ & $m_g^2/m_f^2$ \\ 
\hline 
$1$ & $1$ & $5.08$ & $0.0815$ & $25.8 $ \\
$10^{-12}$ & $1$ & $8.85$ & $5.11\times 10^{-11}$ & $7.84 \times 10^{-11}$ \\ 
$1$ & $10^{-12}$ & $4.00$ & $9.34\times 10^{-14}$ & $16.0$ \\
$10^{-6}$ & $10^{-6}$ & $4.00$ & $8.10\times 10^{-11}$ & $1.60 \times 10^{-5}$ \\
\hline 
 \end{tabular}
\label{heavy}
\end{center}
\end{table}

As a result, 
although the graviton mass square and the cosmological constant
are ordinarily the same as $m_{\rm eff}^2 \sim \Lambda_g$, 
it is possible to find a much heavier graviton mass 
compared with the observed cosmological constant.

\section{Cosmological linear perturbations}
\label{comological_perturbation}
In this Appendix, we shortly summarize 
the linear perturbations of a flat FLRW universe 
 in the bigravity theory\cite{cp_theory}.
Just for simplicity, we assume that the background metrics are 
given by the homothetic flat FLRW spacetimes.
The detail analysis for more generic background 
spacetime including vector and tensor modes 
was discussed in \cite{cp_instability2}.

The background homothetic flat FLRW spacetimes are given by
\begin{align}
\gB_{\mu\nu}dx^\mu dx^\nu &=-dt^2+a^2(t)\delta_{ij}dx^idx^j\,,\\
\fB_{\mu\nu}&=K^2\gB_{\mu\nu}\,.
\end{align}
This background solution is determined by the standard Friedmann equation
with a cosmological constant
 and the following constraints must be satisfied:
\begin{align}
\kappa_g^2\TB{}^{[\rm m]\mu}{}_{\nu}
&=K^2\kappa_f^2\cTB{}^{[\rm m]\mu}{}_{\nu}\,,\\
\Lambda_g&=K^2\Lambda_f\,.
\end{align}

We then consider the adiabatic scalar perturbations and 
ignore an anisotropic stress.
The perturbed metrics 
 are expressed as 
\begin{align}
g_{00}&=-(1+2\alpha_g Y)\,,\nn
g_{0i}&= -a\beta_g^{(L)} Y_i,\nn
g_{ij}&=a^2(\delta_{ij}+2h_g^{(L)}\delta_{ij}Y+2h_g^{(T)}Y_{ij})\,, \\
f_{00}&=-K^2(1+2\alpha_f Y)\,,\nn
f_{0i}&= -K^2a\beta_f^{(L)} Y_i\,,\nn
f_{ij}&=K^2a^2(\delta_{ij}+2h_f^{(L)}\delta_{ij}Y+2h_f^{(T)}Y_{ij})\,,
\end{align}
while the perturbed energy-momentum tensors are given by
\begin{align}
T^0{}_0&=-\bar{\rho}_g(1+\delta_g)\,,\nn
T^0{}_i&=a(\bar{\rho}_g+\bar{P}_g)(v^{(L)}_g-\beta^{(L)}_g)Y_i\,,\nn
T^i{}_0&=-a^{-1}(\bar{\rho}_g+\bar{P}_g)v^{(L)}_g Y^i \,, \nn
T^i{}_j&=P_g(\delta^i{}_j+\pi_g^{(L)}\delta^i{}_j)
\,,\\
\mathcal{T}^0{}_0&=-\bar{\rho}_f(1+\delta_f)\,,\nn
\mathcal{T}^0{}_i&=a(\bar{\rho}_f+\bar{P}_f)(v^{(L)}_f-\beta^{(L)}_f)Y_i\,,\nn
\mathcal{T}^i{}_0&=-a^{-1}(\bar{\rho}_f+\bar{P}_f)v^{(L)}_f Y^i \,, \nn
\mathcal{T}^i{}_j&=P_f(\delta^i{}_j+\pi_f^{(L)}\delta^i{}_j)\,,
\end{align}
where 
the scalar harmonic function $Y$ is defined by
\begin{align}
(\Delta +k^2)Y=0\,,
\end{align}
with $-k^2$ being an eigenvalue of the usual 
three-dimensional Laplacian operator $\Delta$,
and  its vector and tensor harmonic functions are defined by:
\begin{align}
Y_i&=-k^{-1}\partial_i Y \,, \nn
Y_{ij}&=k^{-2}\left( \partial_i\partial_j
-\frac{1}{3}\delta_{ij}\partial^a\partial_a\right)Y
\,,
\end{align}
respectively.
The perturbation variables 
$\{ \alpha_{g/f}, \beta_{g/f}^{(L)}, h_{g/f}^{(L)}, h_{g/f}^{(T)}\}$ and 
$\{ \delta_{g/f}, v^{(L)}_{g/f}, \pi_{g/f}^{(L)} \}$
depend only on time.
The unperturbed energy densities
and pressures, 
$\{ \bar{\rho}_{g/f}, \bar{P}_{g/f} \}$, 
must satisfy 
\begin{align}
\kappa_g^2\bar{\rho}_g=K^2\kappa_f^2\bar{\rho}_f\,,
\quad 
\kappa_g^2\bar{P}_g=K^2\kappa_f^2\bar{P}_f\,.
\label{matter_prop}
\end{align}

For the perturbation variables in the $g$-world,
we can define the gauge invariant variables as in GR:
\begin{align}
\Phi_g&=\alpha_g-\dot{\sigma}^{(L)}_g, \nonumber \\
\Psi_g&=\mathcal{R}_g-H\sigma^{(L)}_g, \nonumber \\
\Delta_g&=\delta_g+3(1+w)\frac{a}{k}H(\beta^{(L)}_g-v^{(L)}_g)_g, \nonumber \\
V_g&=v^{(L)}_g+\frac{a}{k}\dot{h}^{(T)}_g, 
\end{align}
where
\begin{align}
w=\bar{P}_g/\bar{\rho}_g
,\quad
c_s^2=\dot{\bar{P}}_g/\dot{\bar{\rho}}_g \,.
\end{align}
$\mathcal{R}_g$ and $\sigma_g$ are the curvature 
and the shear perturbations, respectively,
which are defined by
\begin{align}
\mathcal{R}_g&= h_g^{(L)}+\frac{1}{3}h_g^{(T)} \,,\\
\sigma^{(L)}_g &=\frac{a^2}{k^2}\dot{h}^{(T)}_g-\frac{a}{k}\beta^{(L)}_g \,.
\end{align}

Similarly, we introduce the gauge invariant variables in the $f$-world, 
which are defined by those with the subscript $f$.
We note $w$ and $c_s^2$ coincide in the $g$- and $f$-worlds
because of \eqref{matter_prop}.

The massless  and massive mode perturbations, $X_+$ and $X_-$,
are described by the linear combination of the perturbed variables
in the $g$- and $f$-worlds,  $X_{g}$ and $X_{f}$,  as
\begin{align}
X_+&=\frac{m_f^2}{m_{\rm eff}^2}X_g+\frac{m_g^2}{m_{\rm eff}^2}X_f\,, \\
X_-&=X_g-X_f\,,
\label{+-gf}
\end{align}
or inversely
\begin{align}
X_g&=X_++\frac{m_g^2}{m_{\rm eff}^2}X_-\,, \\
X_f&=X_+-\frac{m_f^2}{m_{\rm eff}^2}X_-\,.
\label{gf+-}
\end{align}

For the massless mode, there are four independent  equations
\begin{align}
-\frac{k^2}{a^2}\Phi_+
&=\frac{\kappa_g^2\bar{\rho}_g}{2}\Delta_+ \\
\Phi_+ +\Psi_+
&=0 \,,
\\
\dot{\Delta}_+ -3wH\Delta_+ +(1+w)\frac{k}{a}V_+
&=0, 
\end{align}
and
\begin{align}
\dot{V}_+ +HV_+ -\frac{k}{a}\left[\frac{c_s^2\Delta_+}{1+w}+\Phi_+ \right]
&=0\,,
\end{align} 
for four perturbation variables $\{\Phi_+,\Psi_+,\Delta_+,V_+\}$.

If both background matter densities ($\bar\rho_g$ and $\bar\rho_f$) are dominated by 
non-relativistic matter $(w=0)$, 
the equation for the density perturbation $\Delta_+$  is given by
\begin{align}
\ddot{\Delta}_+ +2H\dot{\Delta}_+
+\left(\frac{k^2c_s^2}{a^2}-\frac{\kappa_g^2\rho_g}{2}\right)\Delta_+=0\,,
\label{density_perturbation_massless}
\end{align}
which is the same as that in GR.
Then we will not discuss it furthermore.

Unlike the massless mode, 
there are six independent equations of motion for the massive mode 
variables 
$\{ \alpha_{-}, \beta_{-}^{(L)}, h_{-}^{(L)}, h_{-}^{(T)}, \delta_{-}, v^{(L)}_{-}
 \}$.
By use of $ \Phi_-, \Psi_-$, which are given by the above six variable,
we find the similar four equations to those of the massless mode as
\begin{gather}
-\frac{k^2}{a^2}\Phi_- +m_{\rm eff}^2
\left(\frac{3}{2}h^{(L)}_-
+\frac{3}{4}\frac{a}{k}H\beta^{(L)}_- +h^{(T)}_-\right) =\frac{\kappa_g^2\bar{\rho}_g}{2}\Delta_-, 
\label{cp_eq1}\\ 
\Phi_- +\Psi_- =
m_{\rm eff}^2\frac{a^2}{k^2}h^{(T)}_-,
\label{cp_eq2} \\
\dot{\Delta}_- 
-3wH\Delta_- 
+(1+w)\frac{k}{a}V_- \nn
+\frac{3}{4}(1+w) m_{\rm eff}^2 \frac{a}{k}\beta^{(L)}_-=0,
\label{cp_eq4}\\
\dot{V}_- +HV_- 
-\frac{k}{a}
\left[\frac{c_s^2\Delta_-}{1+w}+\Phi_-\right]=0,
\label{cp_eq5}
\end{gather}
in which the extra terms come from the interactions 
between two metrics.
In addition, we have two more independent equations 
from \eqref{pert_varphi-Bianchi} as
\begin{align}
&6\dot{h}^{(L)}_- +6H h^{(L)}_- -6H\alpha_- +\frac{k}{a}\beta^{(L)}_-=0\,,
\label{cp_eq6}\\
&\frac{a}{k}\left( \frac{3}{2}\dot{\beta}^{(L)}_- 
+6H\beta^{(L)}_- \right)+3\alpha_- +6h^{(L)}_- +2h^{(T)}_-=0\,.
\label{cp_eq7}
\end{align}
Note that although the massive mode variables are gauge invariant in themselves,
we also use $\Phi_-,\Psi_-, \Delta_-$ and $V_-$ just for 
the similar description to those of the massless mode.

Once the equation of state are given, since the above six dynamical 
equations are independent,
we can solve the six variables 
$\{ \alpha_{-}, \beta_{-}^{(L)}, h_{-}^{(L)}, h_{-}^{(T)}, \delta_{-}, v^{(L)}_{-}\}$
for given appropriate initial data.

In order to set up initial data, 
we have the additional constraint equations:
\begin{align}
&(3m_{\rm eff}^2-2\Lambda_g)\left(\alpha_- +3h^{(L)}_-\right)
\nn
&=
\kappa_g^2\bar{\rho}_g \left( \delta_- -3w\pi^{(L)}_-
-(1+3w)\alpha_- +3(1-w)h^{(L)}_- \right)\,,
\label{cp_trace_eq} \\
&-H\Phi_- +\dot{\Psi}_-
=\frac{a}{k}\dot{H}V_- 
+\frac{1}{4}m_{\rm eff}^2\frac{a}{k}\beta^{(L)}_- \,.
\end{align}
which are obtained from \eqref{pert_varphi-trace} and
$0$-$i$ component of the Einstein equations.

From the above basic equations, we find that the variables 
consist of two parts: One is an oscillating wave part and the other 
is a monotonically changing part in time. 
As an example, we show the equation for $h_-^{(T)}$:
\bea
&&
\ddot h_-^{(T)}+3 H \dot h_-^{(T)}+\left(
{k^2\over a^2}+m_{\rm eff}^2\right)h_-^{(T)}
\nn
&&
=
-{k^2\over a^2}\left(
\alpha_-+3h_-^{(L)}\right)+12H\left(
\dot h_-^{(L)}+Hh_-^{(L)}-H\alpha_-\right)
\nn
&&
\approx
-{k^2\over a^2}\left(
\alpha_-+3h_-^{(L)}\right)~~\left({\rm for} ~~
{a\over k}\ll H^{-1} \right)
\,.
\ena

This equation naively shows that 
$h_-^{(T)}$ oscillates with the frequency $\omega\sim 
\sqrt{{k^2/a^2}+m_{\rm eff}^2}$ with the damping amplitude 
due to the expansion of the universe ($H$). 
Although the right hand side may work as a source term, 
which could increase the amplitude, 
it is not the case as we show it numerically.
As a result, the metric variable will approach 
a monotonically changing part with damping oscillations.


\begin{thebibliography}{99}
\bibitem{footnote1}
   Although we focus on the ghost-free bigravity theory in this paper, 
many bigravity theories have been proposed 
\cite{bigravity1,bigravity2,bigravity3,bigravity4,bigravity5,bigravity6,bigravity7},
for which we may also find the similar results.
\bibitem{footnote2}
Some people call it the quasi-static approximation \cite{cp_instability3}.
\end{thebibliography}
\end{document}